\documentclass{article}

\usepackage{arxiv}

\usepackage[utf8]{inputenc} 
\usepackage[T1]{fontenc}    
\usepackage{hyperref}       
\usepackage{url}            
\usepackage{booktabs}       
\usepackage{amsfonts}       
\usepackage{nicefrac}       
\usepackage{microtype}      
\usepackage{lipsum}		
\usepackage{graphicx}
\usepackage{natbib}
\usepackage{doi}

\usepackage{titlesec} 
\usepackage{mathtools}
\usepackage{breqn}

\title{Additive manufacturing process design with differentiable simulations}

\author{ \hspace{1mm}Mojtaba Mozaffar \\
	Department of Mechanical Engineering\\
	Northwestern University\\
	Evanston, IL 60208, USA \\
	\texttt{mozaffar@u.northwestern.edu}
	\And
	\hspace{1mm}Jian Cao\thanks{Corresponding author} \\
	Department of Mechanical Engineering\\
	Northwestern University\\
	Evanston, IL 60208, USA \\
	\texttt{jcao@northwestern.edu}
}

\renewcommand{\headeright}{}
\renewcommand{\shorttitle}{Additive manufacturing process design with differentiable simulations}

\hypersetup{
pdftitle={Additive manufacturing process design with differentiable simulations},
pdfsubject={},
pdfauthor={Mojtaba Mozaffar, Jian Cao},
pdfkeywords={Additive Manufacturing, Design, Differentiable Simulation, Automatic Differentiation},
}

\titleformat{\subsubsection}[runin]
  {\normalfont\normalsize\bfseries}{\thesubsubsection}{1em}{}

\begin{document}
\maketitle

\begin{abstract}
We present a novel computational paradigm for process design in manufacturing processes that incorporates simulation responses to optimize manufacturing process parameters in high-dimensional temporal and spatial design spaces. We developed a differentiable finite element analysis framework using automatic differentiation which allows accurate optimization of challenging process parameters such as time-series laser power. We demonstrate the capability of our proposed method through three illustrative case studies in additive manufacturing for: (i) material and process parameter inference using partial observable data, (ii) controlling time-series thermal behavior, and (iii) stabilizing melt pool depth. This research opens new avenues for high-dimensional manufacturing design using solid mechanics simulation tools such as finite element methods. Our codes are made publicly available for the research community at https://github.com/mojtabamozaffar/differentiable-simulation-am.
\end{abstract}


\section{Introduction}

While pure data-driven modeling approaches can offer computational efficiency and flexibility in many applications, they often introduce errors in predicting challenging scenarios outside of their training domain. On contrary, physics-based modeling can conserve known physical laws over arbitrary domains. Due to the difference in weaknesses and strengths of these two approaches, one can imagine that a combination of the two, i.e., a physics-informed data-driven model, can lead to superior performance. One way to achieve such a hybrid model is to decompose the simulation into subtasks, some of which are solved using physics-based simulations while using a data-driven approach for others. For example, a physics-based model can be deployed when a subtask of simulation involves a trustworthy known physics with affordable computational cost exists, whereas parts with unreliable physics or conventionally expensive simulations can be replaced with data-driven modeling. As most modern data-driven approaches involve gradient-based optimization, to optimize hybrid simulations both physics- and data-driven-based tasks need to be differentiable, i.e., allow calculation of the gradient of their outputs with respect to their inputs and internal variables. In addition to that, differentiable physics-based simulations have many stand-alone applications in scientific computing as they provide instantaneous access to high-dimensional gradients which are necessary for most solvers and optimizers.

In recent years, differentiable simulations are increasingly used in robotics to advance the modeling and control capabilities. Hu et al. \citep{1} used a differentiable kinematic model in a model-based reinforcement learning setting for soft robots. Heiden et al. \citep{2} devised a differentiable simulation of robotic rigid body motion which led to an accurate system identification model with visual inputs. Other noteworthy publications in this field include \citep{3,4,5}; however, the state-of-the-art studies are limited to robotics and particle-based systems.

\begin{figure}
	\centering
	\includegraphics[width=14cm]{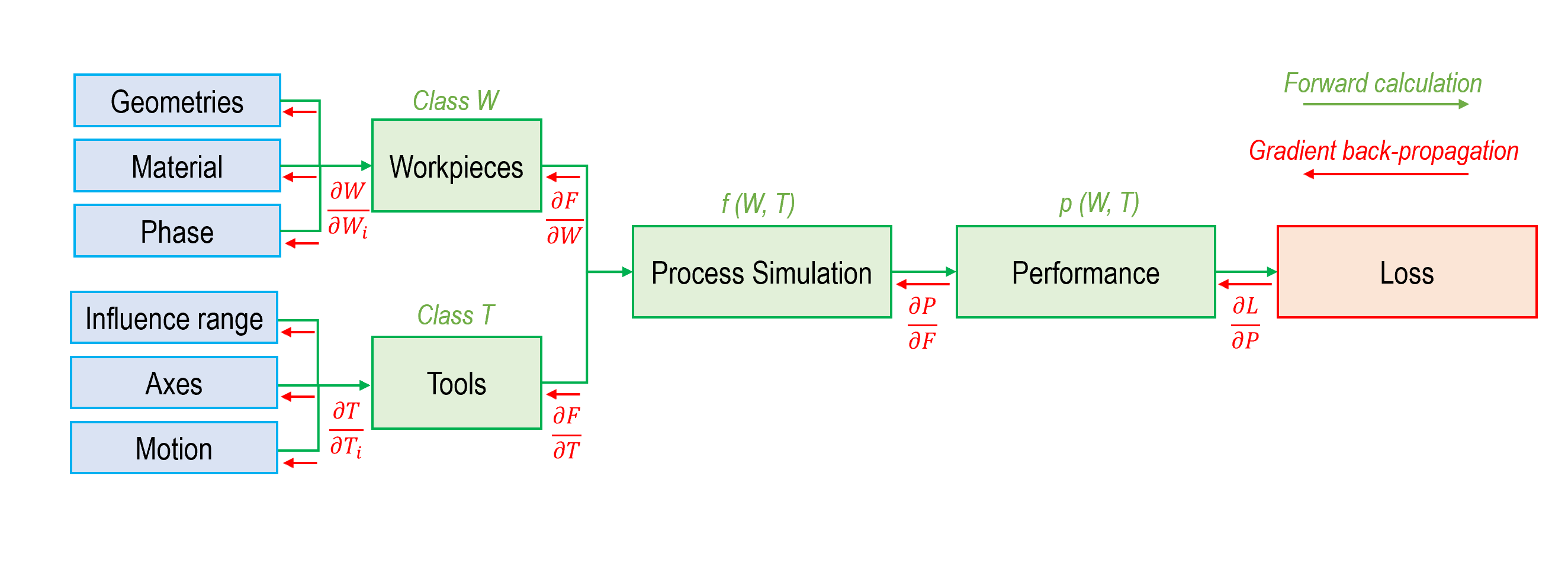}
	\caption{Differentiable manufacturing process simulation capable of calculating the gradients of performance loss with respect to workpiece, tool, and process parameters.}
	\label{fig:fig1}
\end{figure}

In this article, we present a differentiable computational paradigm for process design in manufacturing processes that incorporates differentiable physics-based simulation and data-driven responses to optimize manufacturing process parameters in high-dimensional temporal and spatial design spaces. In particular, we aim to answer two key questions: (i) can the gradients of desired build performance be efficiently computed in manufacturing processes, and (ii) would gradient-based optimization provide an effective tool to optimize manufacturing processes in challenging environments. In a general manufacturing process, the simulation tool determines the interaction of a set of workpieces (including start and target geometry, material, etc.) and manufacturing tools as demonstrated in Fig.\ref{fig:fig1}. Given the initial process parameters, one can compute the performance of the physics-based simulation model by going through a forward pass (green arrows in Fig.\ref{fig:fig1}) of the computational scheme. Using a differentiable simulation, the gradients of a loss function (defined based on the desired performance) with respect to any of the workpiece, tools, or process parameters can be computed (red arrows in Fig.\ref{fig:fig1}), which can be used in an optimization setting to design manufacturing inputs.

In what follows, we first introduce the background on automatic differentiable in Section 2. We provide details of our methodology for differentiable finite element simulations in Section 3 and demonstrate the capability of our proposed approach through three illustrative case studies in Section 4. Finally, we conclude this paper by summarizing our findings and laying down our vision for future research topics in Section 5.

\section{Automatic Differentiation and Libraries}

The surge in the field of artificial intelligence and neural network predictive modeling is partially due to heterogeneous high-performance computing capabilities and graph-based automatic differentiation, which enables us to calculate the gradients of an arbitrary loss function with respect to any of the internal weights in a neural network and, therefore, efficiently navigate through high-dimensional weight spaces. A core idea in this research is to develop the computational graph for a physics-based simulation of manufacturing and utilize the gradients of various high-dimensional process parameters with respect to the desired performance to come up with novel design solutions.

In an automatic differentiation scheme, we construct the computational process as a composition of operations, where the gradient of each operation is known. Each operation represents a directed node in the computational graph. The forward pass computes the outputs of the simulation given the inputs while storing intermediate results. The backward pass starts from the output node and recursively computes the gradient of parameters by multiplying the incoming gradient from the next node and the partial derivation of the node evaluated at the current value. More generally, the gradient between any two parameters on the same computational graphs can be calculated by (1) finding a computational path between parameters and (2) multiplying the gradient contribution of each operation along the way. Note that such a computational path is unique by construction.

\begin{figure}
	\centering
	\includegraphics[width=10cm]{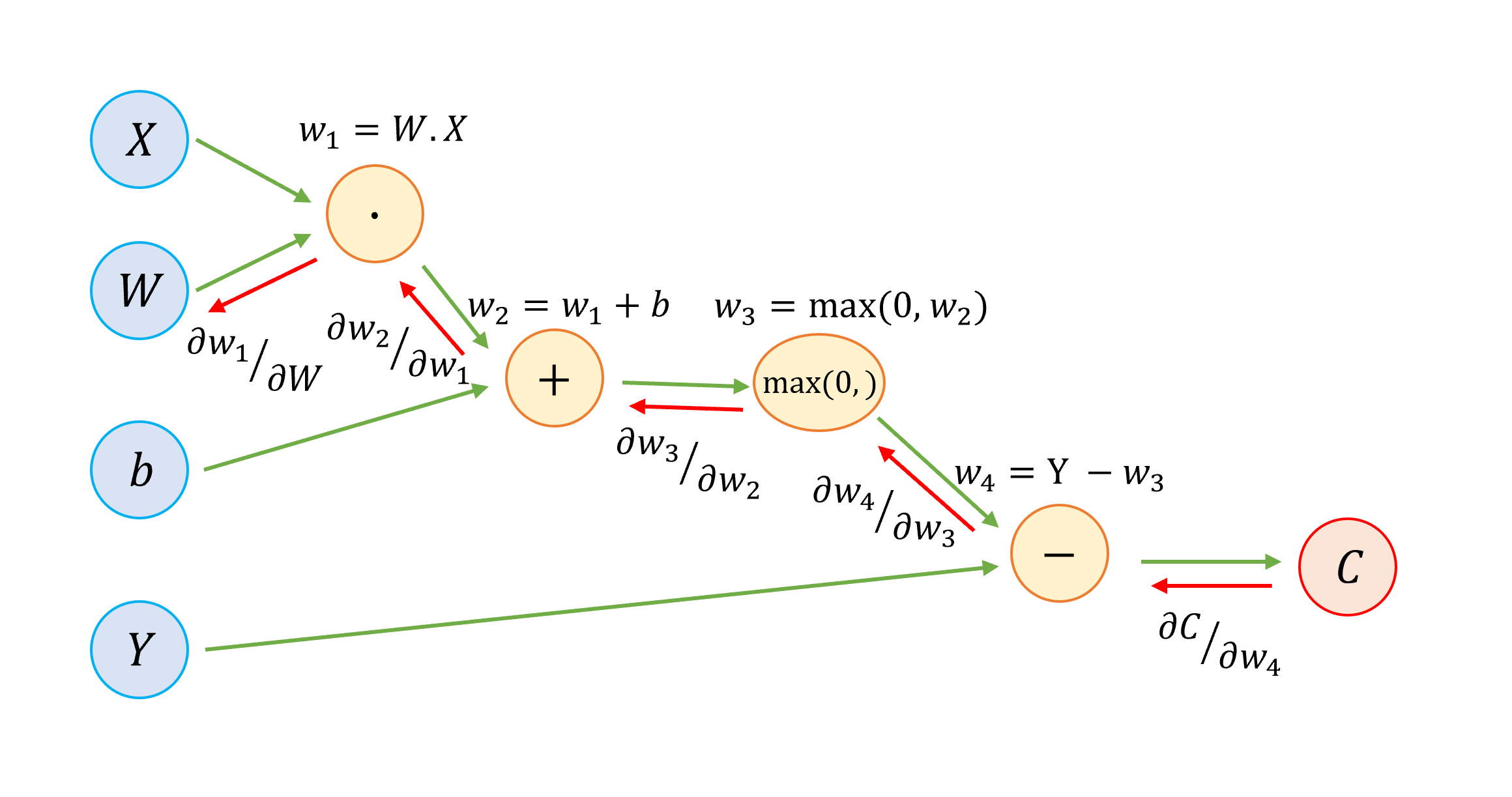}
	\caption{Schematic of a computational graph for computing the cost function ($C$) in $C = Y - tanh(W.X + b)$. This computational graph can be utilized the forward calculation of cost function as well as backpropagation calculation of gradients.}
	\label{fig:fig2}
\end{figure}

As a simple example, consider the computation of the loss function in a regression task with the following formula: 

\begin{equation} \label{eq:1}
C = Y - tanh(W.X + b)
\end{equation}

where $X$ is the input, $W$ and $b$ are trainable weight and bias parameters, $tanh$ is the hyperbolic tangent function, and $Y$ is the correct label. A computational graph of this computation can be constructed as demonstrated in Fig.\ref{fig:fig2}, where operations stem from input nodes in blue and gradually build toward the cost function.

Assuming initial values of $X=1$, $W=2$, $b=3$, and $Y=6$, we can sequentially compute and store the intermediate variables and the cost as $w_1=2$, $w_2=5$, $w_3=5$, $w_4=1$, and $C=1$. For computing the gradients of the cost with respect to the weight, $\frac{\partial C}{\partial W}$ , we can find a path between these two graph nodes. Starting from the last node ($C$), we can compute the gradient of the cost with respect to all the nodes on the path. Performing this operation in the reverse orders allows effective use of dynamic programming where the gradient of each node only depends on the upstream gradient (which is readily available due to the reverse order of calculations) and local gradient of that node (which only depends on a known partial derivative function and the value of the node calculated during the forward path). Thus, the gradients can be computed as shown in Table \ref{tab:table1}.

\begin{table}
	\caption{Backpropagation steps for gradient calculations.}
	\centering
	\begin{tabular}{lll}
		\toprule
		$\frac{\partial C}{\partial w_4}=1$   \\
		\\
		$\frac{\partial C}{\partial w_3}= \frac{\partial C}{\partial w_4} * \frac{\partial w_4}{\partial w_3} = 1 * -1 = -1$    \\
		\\
		$\frac{\partial C}{\partial w_2}= \frac{\partial C}{\partial w_3} * \frac{\partial w_3}{\partial w_2} = -1 * 1 = -1$    \\
		\\
		$\frac{\partial C}{\partial w_1}= \frac{\partial C}{\partial w_2} * \frac{\partial w_2}{\partial w_1} = -1 * 1 = -1$    \\
		\\
		$\frac{\partial C}{\partial W}= \frac{\partial C}{\partial w_1} * \frac{\partial w_1}{\partial W} = -1 * X = -1$    \\
		
		\bottomrule
	\end{tabular}
	\label{tab:table1}
\end{table}

Several libraries efficiently construct and compute computational graphs for automatic differentiation in both forward and backward passes, such as Theano \citep{6}, Torch \citep{7}, TensorFlow \citep{8}, PyTorch \citep{9}, JAX \citep{10}, and Taichi \citep{11}. Each of these libraries offers different architecture choices, capabilities, compatibility ecosystem, and performance in different applications.

\section{Methodology}

As a representative manufacturing process physics-based simulation, we select additive manufacturing (AM) thermal simulation to investigate the capabilities, flexibility, and limitations of differentiable simulations for manufacturing design. As mentioned before, the thermal profile of AM processes is a pivotal characteristic of this class of manufacturing processes as it determines microstructural evolution and geometric accuracy. We use a finite element formulation to solve transient heat transfer equations over the simulation domain.

\begin{figure}
	\centering
	\includegraphics[width=14cm]{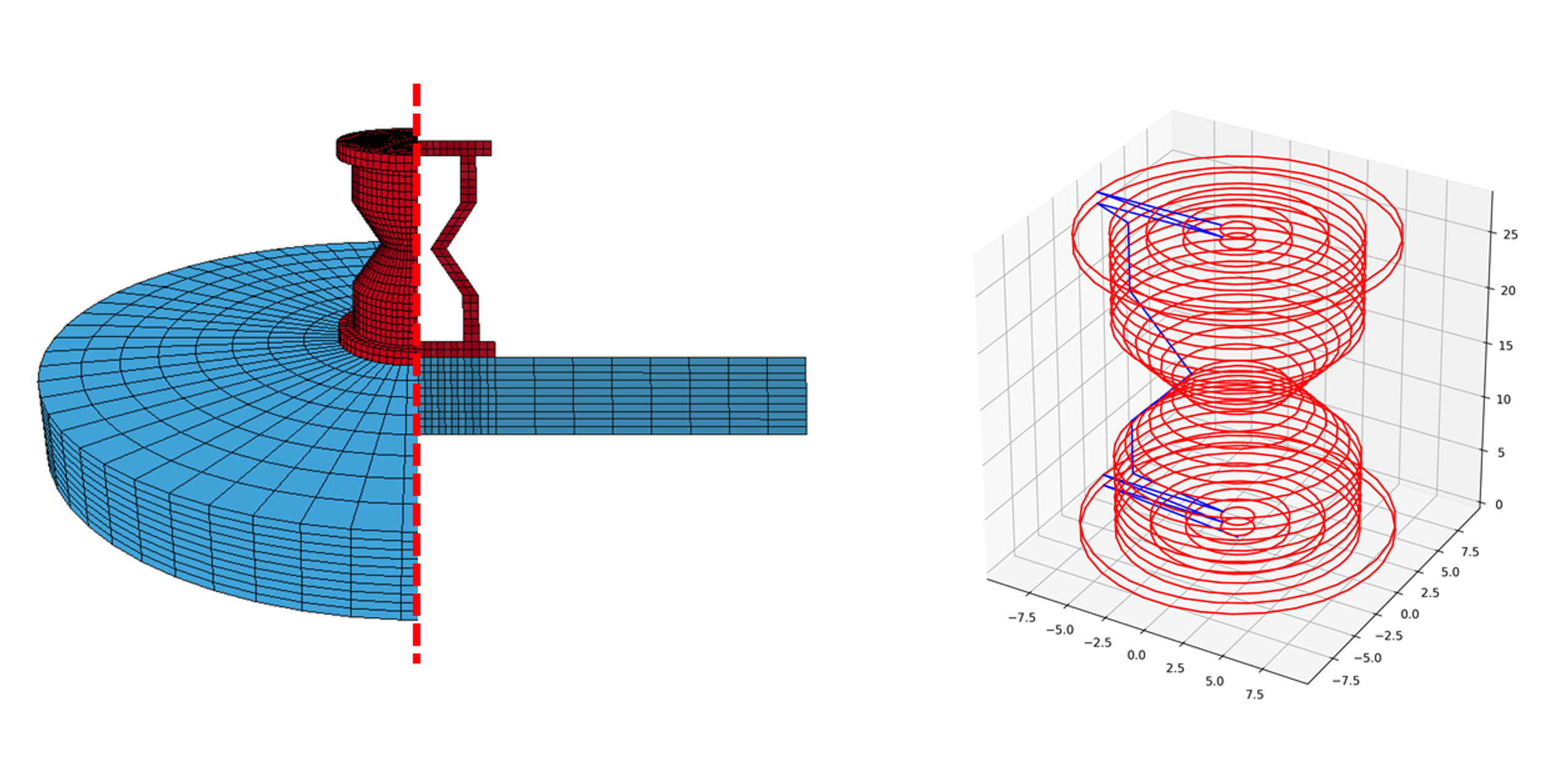}
	\caption{Test case geometry and its cross-section view where red elements represent the build and blue elements are the substrate (left) and toolpath pattern (right) for the differentiable AM. thermal simulations test case. The red lines on toolpath plot indicate nozzle moves while laser is on, while the blue lines indicate motion when laser is off.}
	\label{fig:fig3}
\end{figure}

In this analysis, we first define the geometry through CAD software and produce hexagonal unstructured meshes. While we implemented this method for hexagonal mesh structures, this is merely an implementation choice, and the formulation is capable of capturing other mesh structures such as tetrahedral and higher-order approximations as well. After the mesh, to generate the toolpath, we developed a Python script that slices the CAD geometry at predefined intervals in the vertical direction and produces a toolpath for each 2-dimensional section using a handful of hard-coded strategies, e.g., moving inward from boundaries, zig-zag strategy. We considered an hourglass part as a testbed here. You can see the geometry and generated toolpath for it in Fig.\ref{fig:fig3}. We process the geometry and toolpath to generate an element birth file which indicates the time at which each element would be born in the simulation. These three files (mesh, toolpath, and element birth), along with process parameters including laser characteristics, material properties, and simulation time step will be passed to a differentiable finite element simulation to determine the thermal responses of the AM process.

In the finite element formulation, we aim to solve the partial differential equation (PDE) for transient heat transfer and the boundary conditions over the discretized geometry domain. The heat transfer equations and considered boundary conditions including fixed temperature boundary (i.e., Dirichlet boundary condition), radiation, convection, and laser power flux are provided in Eqs. \ref{eq:2}-\ref{eq:6}.

\begin{equation} \label{eq:2}
\rho c_p  \frac{\partial T}{\partial t} - \nabla . (k . \nabla T) - s = 0
\end{equation}

\begin{equation} \label{eq:3}
(1) Dirichlet \qquad		T=T_1 (x,y,z) \quad on \: \Gamma_1
\end{equation}

\begin{equation} \label{eq:4}
(2) Neumann \qquad		q= -q_s \quad on \: \Gamma_2
\end{equation}

\begin{equation} \label{eq:5}
(3) Neumann \qquad		q= -h(T-T_amb ) \quad on \: \Gamma_3
\end{equation}

\begin{equation} \label{eq:6}
(4) Radiation \qquad		q=-\varepsilon \sigma(T^4-T_{amb}^4 ) \quad on \: \Gamma_4
\end{equation}

where $\rho$ is the material density, $c_p$ is the specific heat capacity, $T$ is temperature, $t$ is time, $k$ is the material conductivity, and $s$ is the heat generate rate per unit volume. $q_s$ is the external heat flux, $h$ is the convection coefficient, $T_{amb}$ is the ambient temperature, $\varepsilon$ is the surface emissivity constant, $\sigma$ is the Stefan-Boltzmann constant, and $\Gamma_1$,  $\Gamma_2$,  $\Gamma_3$, and  $\Gamma_4$ are sets of surfaces that each of these boundary conditions are applied on. Using the aforementioned PDE and the boundary conditions, we drive the finite element weak form and discretize it for each element using shape function ($N^e$) and the derivative of the shape function ($B^e$). Finally, the forward time integration of thermal response can be derived as 

\begin{equation} \label{eq:7}
\{T^{n+1} \}=\{T^n \}+\Delta t[M]^{-1} [\{R_G \}-\{R_F \}-\{R_C \}-\{R_R \}-[K]\{T^n \}]
\end{equation}

where $[M]$ is the capacitance matrix, $[K]$ is the conduction matrix, $\{R_G\}$ is the internal heat vector, $\{R_F \}$ is the external flux vector, $\{R_C \}$  is the convection vector, and $\{R_R \}$  is the radiation vector. $\Delta t$ is the time step and $\{T^{n+1} \}$ and $\{T^n \}$ are nodal temperatures at time steps $n+1$ and $n$, respectively. Each of these global matrix and vectors can be computed by assembling the local contributions of each element to them individually. Interested readers can find a more in-depth derivation of the finite element formulation in \citep{12}. To simulate AM processes, we need to dynamically keep track of active elements and surfaces at each time step and the local contributions of only active objects will be assembled in global matrices.

To use gradient-based optimization methods, it is ideal for all operations to be continuous and differentiable. In addition to that, a careful design is needed when working with operations that can eliminate and saturate the gradient propagation. For instance, a step function stops the propagation of the gradients from a smooth function as its local gradients are zero in all continuous points. As another example, functions such as sigmoid and hyperbolic tangent although generate valid and smooth gradients, if called recursively in the form of $f(…f(f(x)))$ become gradually closer to a step function where they produce gradients very close to zero for all inputs except for a very narrow region where their gradients are very large. This phenomenon is known as the vanishing/exploding gradients and prevents effective optimization using gradient signals. Many sources of discontinuities exist in thermal analysis of AM processes, especially due to the discontinuous nature of material deposition in a meshed domain. Here, we hypothesize that by fixing the geometric and boundary-related discontinuities in the simulation, the main derive behind thermal responses would be the continuous material and process parameters which can be optimized using a differentiable simulation.

As mentioned before, in recent years, many libraries have been developed to create computational graphs and differentiable numerical solutions using automatic differentiation, most of which are focused on the operations common in deep learning such as convolution and various matrix operations. Physics-based simulations inherently require a more diverse set of operations such as random indexing and large-scale atomic operations. Therefore, there is a need to investigate the performance and capability of the existing libraries to perform thermal simulations in a dynamic domain of AM processes.

In this research, we developed four implementations of AM simulators for PyTorch, TensorFlow, JAX, and Taichi libraries and optimized each implementation according to each library’s guidelines. The result of this analysis is summarized in Table \ref{tab:table2}. To the best of our knowledge, TensorFlow version 2.1 lacks the flexible indexing capability required to perform assembly operations in FEM, and therefore, we were not able to develop a successful implementation of differentiable AM simulation using this library. While the other three libraries showed adequate capabilities to build and differentiate through the simulation stack, we observed a substantial gap in the memory consumption and processing time between these three libraries. On a benchmark task, PyTorch used an unreasonable amount of memory (127 GB), JAX operation time is unacceptable with one simulation taking over 10 hours. This is because these libraries offer highly optimized high-level operations in computer vision and natural language processing applications, but large-scale usage of low-level operations, which are ubiquitous in AM simulations, leads to inefficient buffer allocations and GPU kernel launches and severely hurts the overall performance. These performance issues prevent these implementations to be realistically used in even moderately large simulation scenarios.  In our investigation, Taichi library showed a favorable performance with 2X smaller memory usage and 8X smaller processing time compared to the best performance of other libraries as they provide efficient support for GPU mega-kernels, flexible indexing, and atomic operations. Therefore, the Taichi library is used to perform automatic differentiation in the rest of this article.

\begin{table}
	\caption{Performance comparison of prominent automatic differentiation libraries for manufacturing simulations.}
	\centering
	\begin{tabular}{p{0.2\linewidth}|p{0.3\linewidth}|p{0.1\linewidth}|p{0.25\linewidth}}
		\toprule
		Automatic Differentiation Libraries     & Calculation Capability Support for Operations Needed in FEM     & Memory Usage   &   Calculation Time for One Optimization Iteration \\
		\midrule
		PyTorch     & Yes	          & 127 G    &    40 mins  \\
		TensorFlow     & Lack of support for matrix assembly  & --    &  -- \\
		JAX     & Yes	          & 2 G   &  10 hours \\
		Taichi &  Yes            &  1 G      & 5 mins   \\ 
		\bottomrule
	\end{tabular}
	\label{tab:table2}
\end{table}

\section{Optimization Process and Results}

Here, we investigate the capability of the developed differentiable AM simulation to optimize various process parameters and material properties of the process in three case studies. In the first case study, we test our framework to optimize static parameters with partially observable data. The second case study optimizes the entire thermal history of the build by manipulating time-series laser power during the build. Finally, the third case study investigates the capability of our framework to stabilize the melt pool depth as a derived feature from the thermal response by optimizing time-series laser power. The details of case studies and their results are elaborated in the following subsections:

\subsection{Parameter inference based on partial data}

\begin{figure}
	\centering
	\includegraphics[width=14cm]{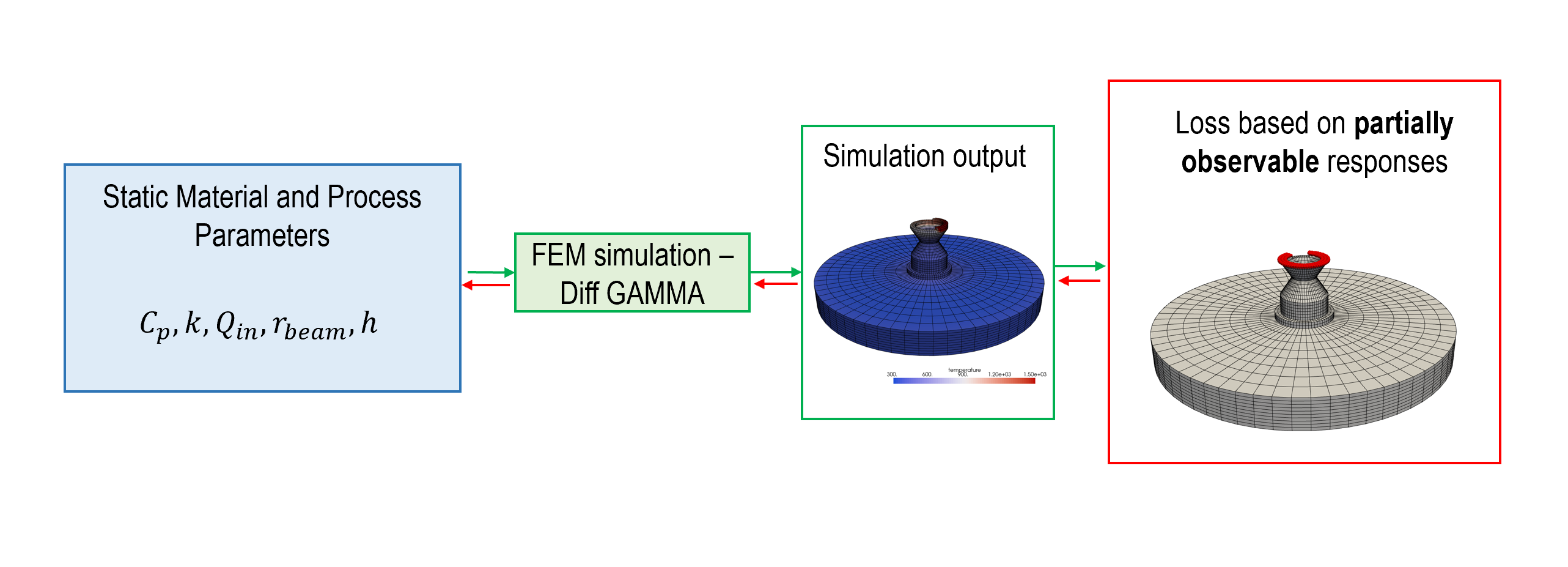}
	\caption{Schematic of the first case study where the partially observable loss function based on the thermal responses of top build layer at each time step is optimized. The optimization paramters include heat capacity, conductivity, convection coefficient, static laser power, and laser beam radius.}
	\label{fig:fig4}
\end{figure}

We devised a case study where we optimize a set of static parameters including material properties and process parameters to obtain a predefined thermal behavior using the build process. The properties investigated in this case include heat capacity, conductivity, convection coefficient, static laser power, and laser beam radius. We initialize the investigated parameters using a uniform distribution over a reasonable range of each parameter. Then, the differentiable simulation generates the output thermal history corresponding to the current parameter set. The loss value is then calculated based on the mean-squared-error (MSE) difference between the target and current responses of the nodes on the top layer of the build at each time step. Finally, the gradient of the loss function with respect to all investigated parameters is calculated using automatic differentiation and the gradient is used to update each parameter using the Adam optimization method \citep{13}. This process is repeated for a set number of iterations, also known as epochs, until we observe a good match between the target and current thermal responses. A schematic of this case study is demonstrated in Fig.\ref{fig:fig4}.

The goal of this analysis is two-fold. First, this case is intended to resemble a model calibration with experimental data, where only part of the build is observable through sensory data such as an IR camera. Therefore, this framework allows us to infer material and process parameters from the process thermal response. Second, the selected set of parameter covers a wide range of operations in FEM analysis, and therefore, this task demonstrates the capability of automatic differentiation to handle many critical operations throughout the constructed computational graph including matrix operations, assembly, lumping, distribution calculation, temporal mapping, to name a few.

\begin{figure}
	\centering
	\includegraphics[width=7cm]{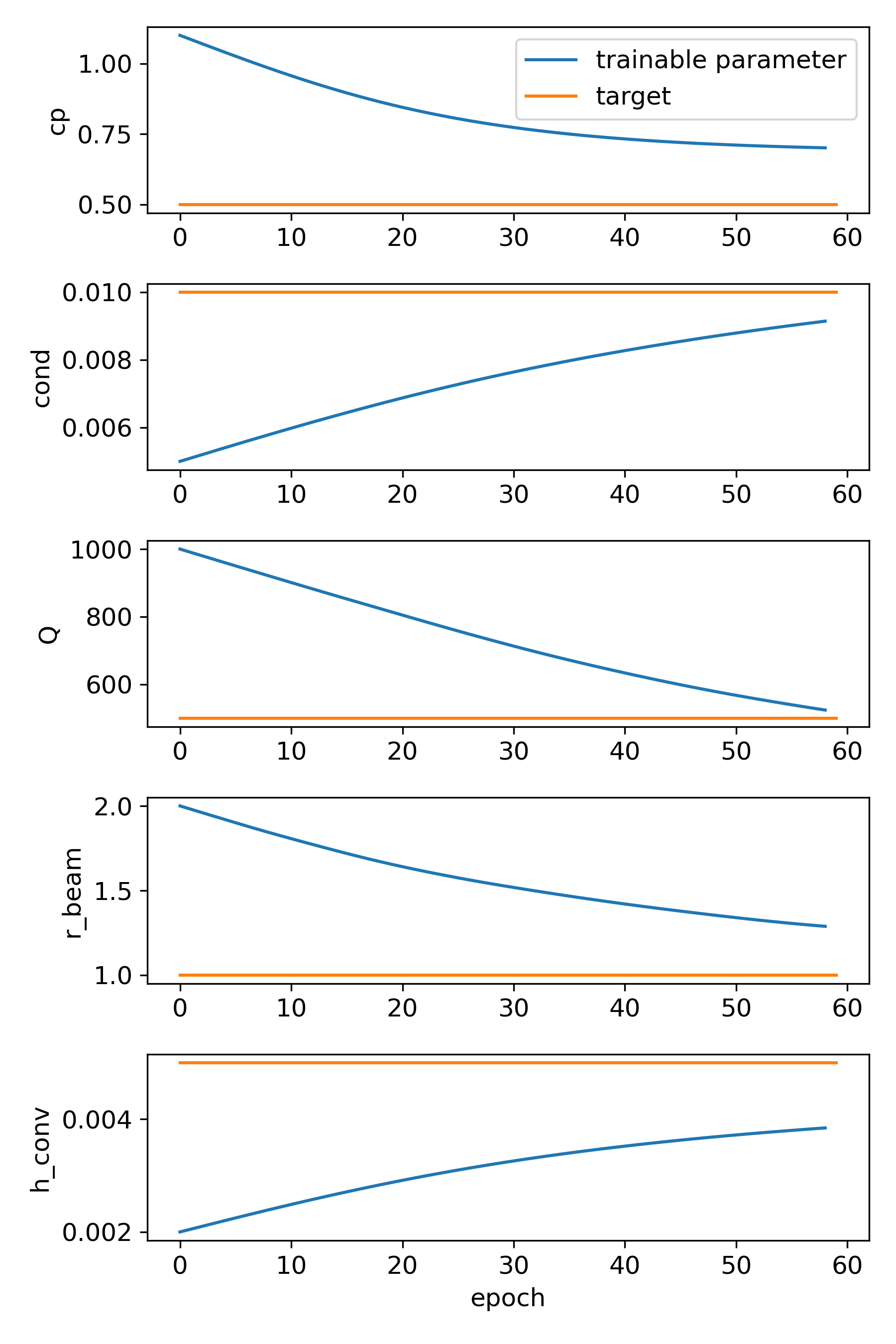}
	\caption{Evolution of the investigated process paramters over $60$ iterations of optimization.}
	\label{fig:fig5}
\end{figure}

The optimization results show that by performing 60 optimization iteration, we reach an error of $1e^{-3}$ MSE on the partially observable loss. Moreover, as shown in Fig.\ref{fig:fig5}, each parameter effectively converges to (or at least moves in the irection of) the parameters that generatedd the target in the loss function. Note that the target parameters in Fig.\ref{fig:fig5} are solely provided as verification of gradient directions and the optimization method does not have access to them; rather, it interprets them using the target thermal response. Overall, this result shows the proposed differentiable method can infer various simulation parameters even when given access to a fraction of simulation responses and elucidates the high potential of differentiable finite element simulations for describing unknown process and simulation parameters. While more optimization steps would bring parameters such as heat capacity closer to their target values, we do not expect it to converge to the exact target as the interaction between optimization parameters and the thermal response is highly coupled. For instance, a similar thermal response can be achieved by underestimating the heat capacity and over-estimating the laser input. Therefore, we believe the fact that all parameters move in the correct direction to collectively reduce the loss to close to zero is a more important result than tuning the optimization steps in a way that each individual parameter reaches its target.

\subsection{High-Dimensional temporal design for thermal history behavior}

\begin{figure}
	\centering
	\includegraphics[width=14cm]{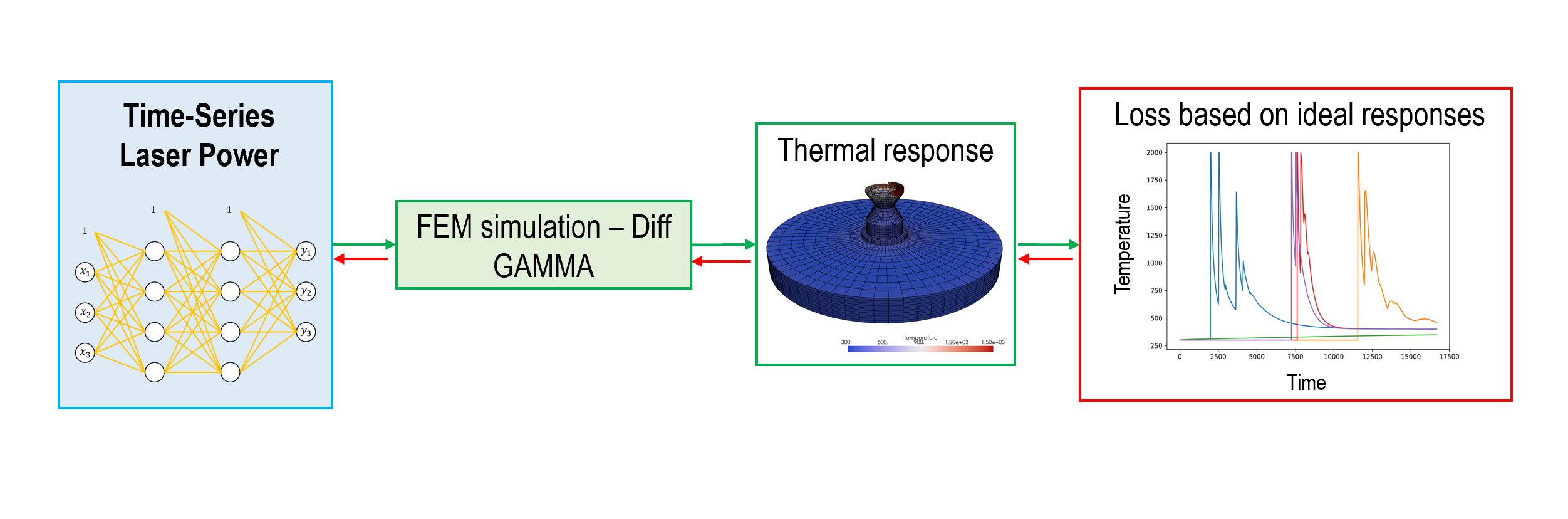}
	\caption{Schematics of the second case study. In this case, a neural network structure determines the time-series laser power of the AM process, and it is optimized to produce an ideal thermal behavior during part build.}
	\label{fig:fig6}
\end{figure}

In the second case study, we explore the capability of differentiable simulation to design high dimensional temporal aspects of the additive manufacturing processes. We use a fully connected neural network architecture to represent the time-series laser power. The neural network receives the time at each time step and outputs the laser power for that time step using two hidden layers of 50 neurons with hyperbolic tangent as the nonlinear activation function to map the output to a range of 0-1,000 W.
We selected this network setting as it allows generating sufficiently complex behavior of laser power over approximately 20,000 time steps of the simulation. Note that these hyperparameters can be adjusted according to the desired nonlinearity in the response. Using this approach, the neural network controls the temporal evolution of the laser power. The computed laser power is then fed into the differentiable simulation and an MSE loss function is defined between the current thermal response and an ideal predefined thermal response. In this case, the ideal thermal profile is developed by simulating the process with a complex laser power pattern. This laser power pattern is not used in the optimization process and is only later used to validate the answer found by differentiable optimization. The optimization task entails computing the gradient of the loss function with respect to weights and biases of neural network and iteratively updating these parameters to minimize the loss. Stainless steel material properties are assigned to the simulation in this case study and unlike the previous case study, they are kept constant during the optimization process. A schematic of this case study is provided in Fig.\ref{fig:fig6}.

\begin{figure}
	\centering
	\includegraphics[width=14cm]{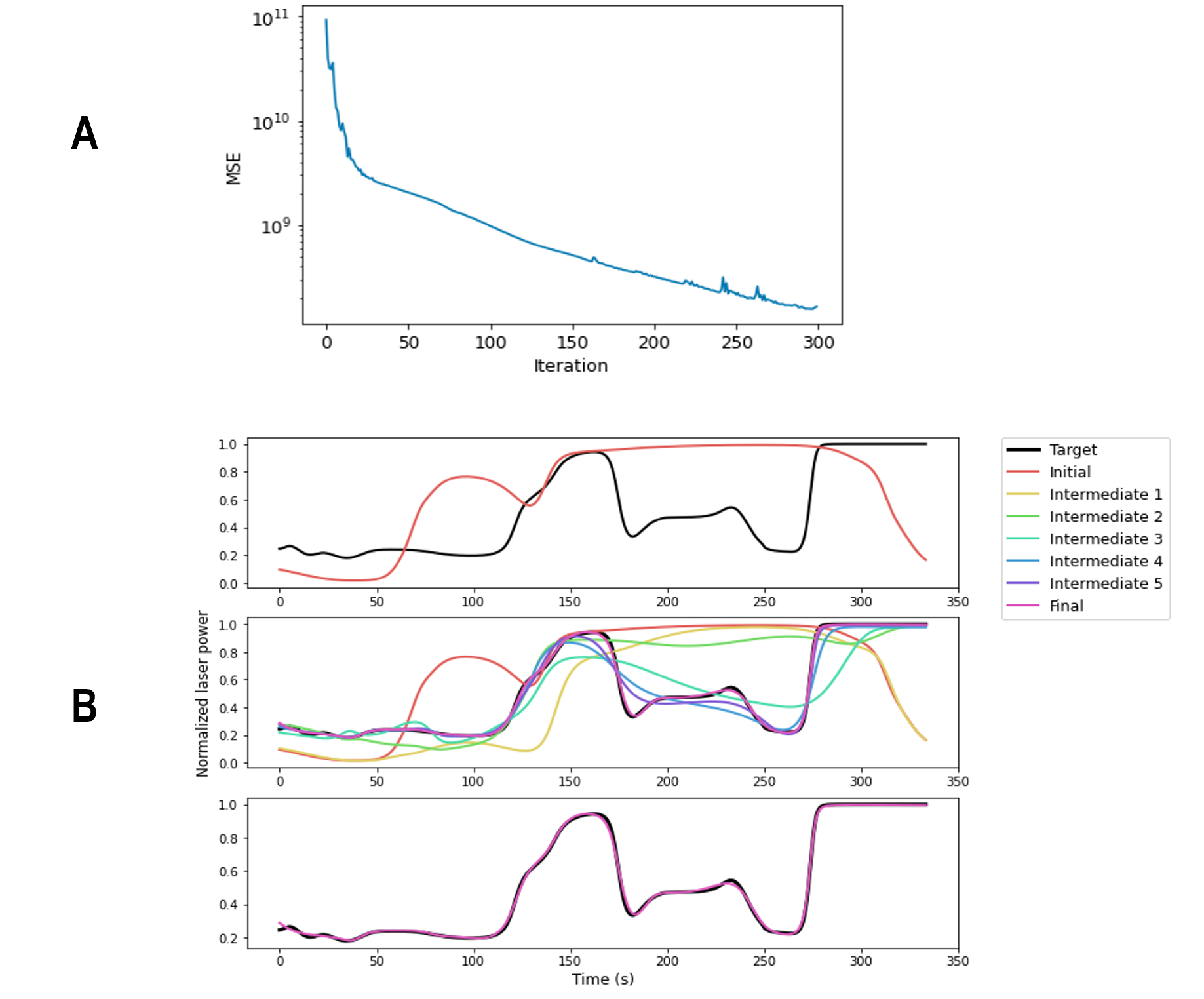}
	\caption{Optimization results for the second case study. (A) evolution of the MSE loss function over 300 optimization iterations. (B) evolution of time-series laser power with the initial laser power plotted in red (see top row), five intermediate laser power patterns during the training (see middle row), and the final pattern found by differentiable optimization after 300 iterations and its comparison with the true target (see bottom row).}
	\label{fig:fig7}
\end{figure}

Similar to the previous case, the Adam algorithm is used to optimize the loss. The evolution of MSE loss over 300 iterations is demonstrated in Fig.\ref{fig:fig7}A which shows the loss decreases about 3 orders of magnitude as the result of optimization. We observe a favorable optimization behavior with the loss function rapidly declining and minimal loss jumps, which shows the suitability of gradient-based methods (albeit with momentum and learning rate scaling) for optimizing time-series parameters in the AM simulation. The evolution of time-series laser power is depicted in Fig.\ref{fig:fig7}B where the true laser power target used to produce the ideal thermal history in the loss function is plotted in black line. This true laser power target is intentionally designed to show sharp changes and complex evolution during the build time. The evolution of the output of the neural network including the initial state, five intermediate states, and the final state after 300 optimization iterations are plotted on the top, middle, and bottom subplots of Fig.\ref{fig:fig7}B correspondingly.

These results indicate that our proposed approach can optimize the time-series values with high accuracy to match an arbitrary target, which is a unique feature of differentiable simulation as it can access accurate gradients of high-dimensional spaces. Additionally, this result exhibits the natural integration of differentiable physics-driven manufacturing simulation with powerful data-driven modeling techniques as another impactful benefit of this approach for the development of physics-informed data-driven methods.

\subsection{High-Dimensional temporal design for melt pool behavior}

\begin{figure}
	\centering
	\includegraphics[width=12cm]{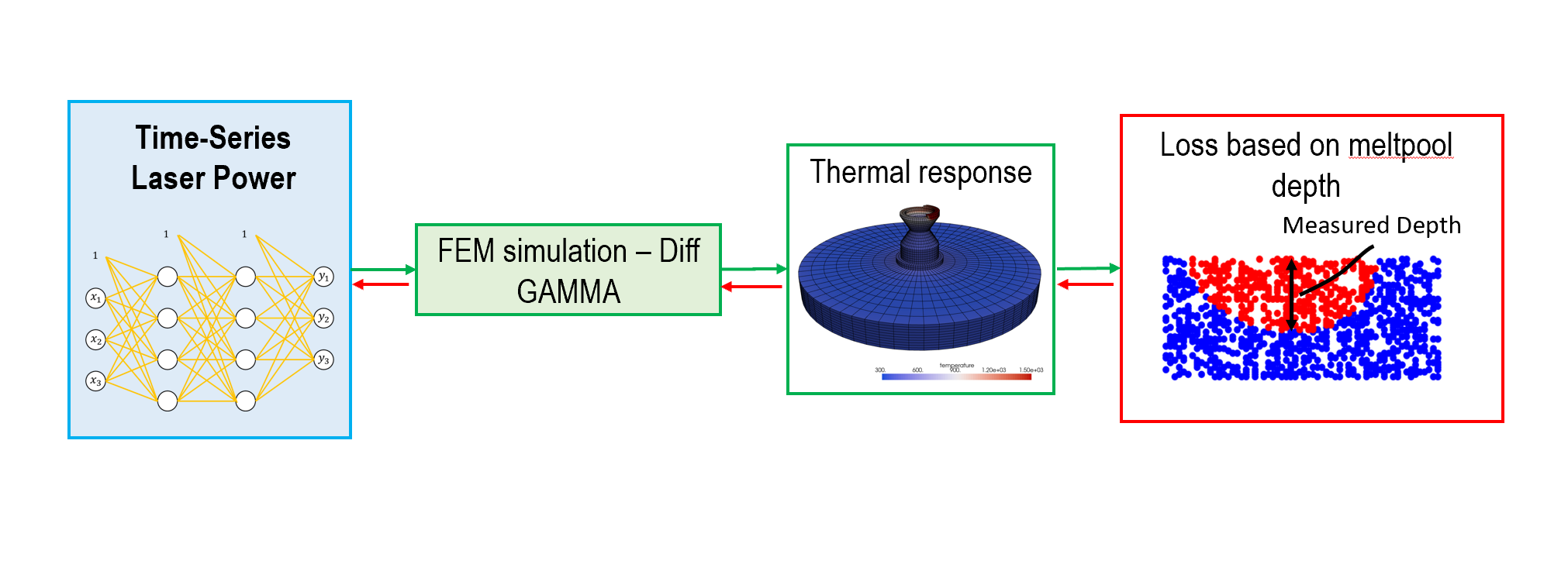}
	\caption{Schematics of the third case study. In this case, we stabilize the melt pool depth throughout the build time by adjusting time-series laser power. The laser power is determines using a fully connected neural network as a universal function approximator and the parameters of the network are tuned using a gradient-based optimization method.}
	\label{fig:fig8}
\end{figure}

In the last two cases, we demonstrated that thermal history can be used as a target for process optimization; however, our proposed computational design approach can be extended to any derivative feature of thermal history that can be computed through a differentiable formulation. In this case, we aim to achieve a target melt pool depth by manipulating time-series laser power. Similar to the second study, we utilize a fully connected neural network to produce time-series laser power that is parameterized by neural network weights and biases. As can be seen from the schematics in Fig.\ref{fig:fig8}, the produced laser power is used in the differentiable AM thermal simulation. Later, the thermal responses are used to calculate the melt pool depth at each time step and an MSE loss function is defined that penalizes the melt pool depth deviations from a predefined depth throughout the build. As melt pool characteristics significantly affect the geometric accuracy of AM processes, investigating systematic solutions to design melt pool features is an important step toward AM parts with customized properties.

\begin{figure}
	\centering
	\includegraphics[width=12cm]{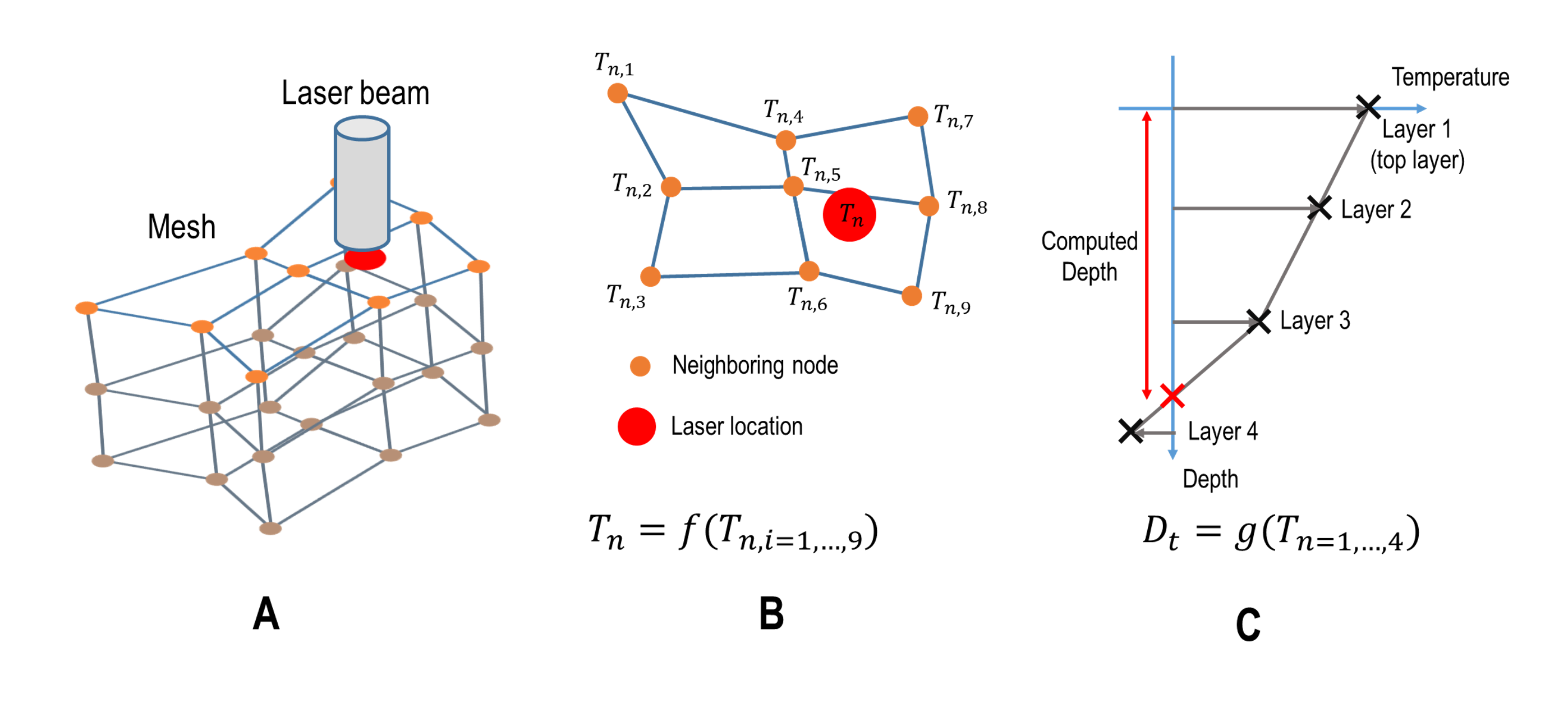}
	\caption{Differentiable melt pool calculation scheme. (A) schematics of a 3 layer mesh structure and the location of laser beam. (B) nodal temperature of nine neighboring nodes are used to compute the temperature corresponding to laser location at each height. (C) a linear pairwise solver is used to compute continuous melt pool depth at each time step.}
	\label{fig:fig9}
\end{figure}

The key to performing this task is to develop a mapping between thermal features and melt pool depth in a way that produces meaningful gradients. For example, calculating the melt pool solely based on the deepest node with a temperature higher than melting temperature although differentiable, does not lead to a helpful optimization method. This is because using the previously mentioned method depth changes similar to a step function which produces zero gradients at each continuous point and therefore stales the gradient-based optimization process. Instead, to compute the continuous representation of the melt pool depth, we dynamically find nine nearest neighboring nodes to the laser location at four height levels starting from the top build layer. At each height level, we interpolate the temperature in the location below the laser beam by solving a ninth-degree system of equations (see Figure Fig.\ref{fig:fig9}B). The temperature bellow laser at each height is then used to compute the continuous melt pool depth using a pairwise linear solver (see Figure Fig.\ref{fig:fig9}C).

\begin{figure}
	\centering
	\includegraphics[width=10cm]{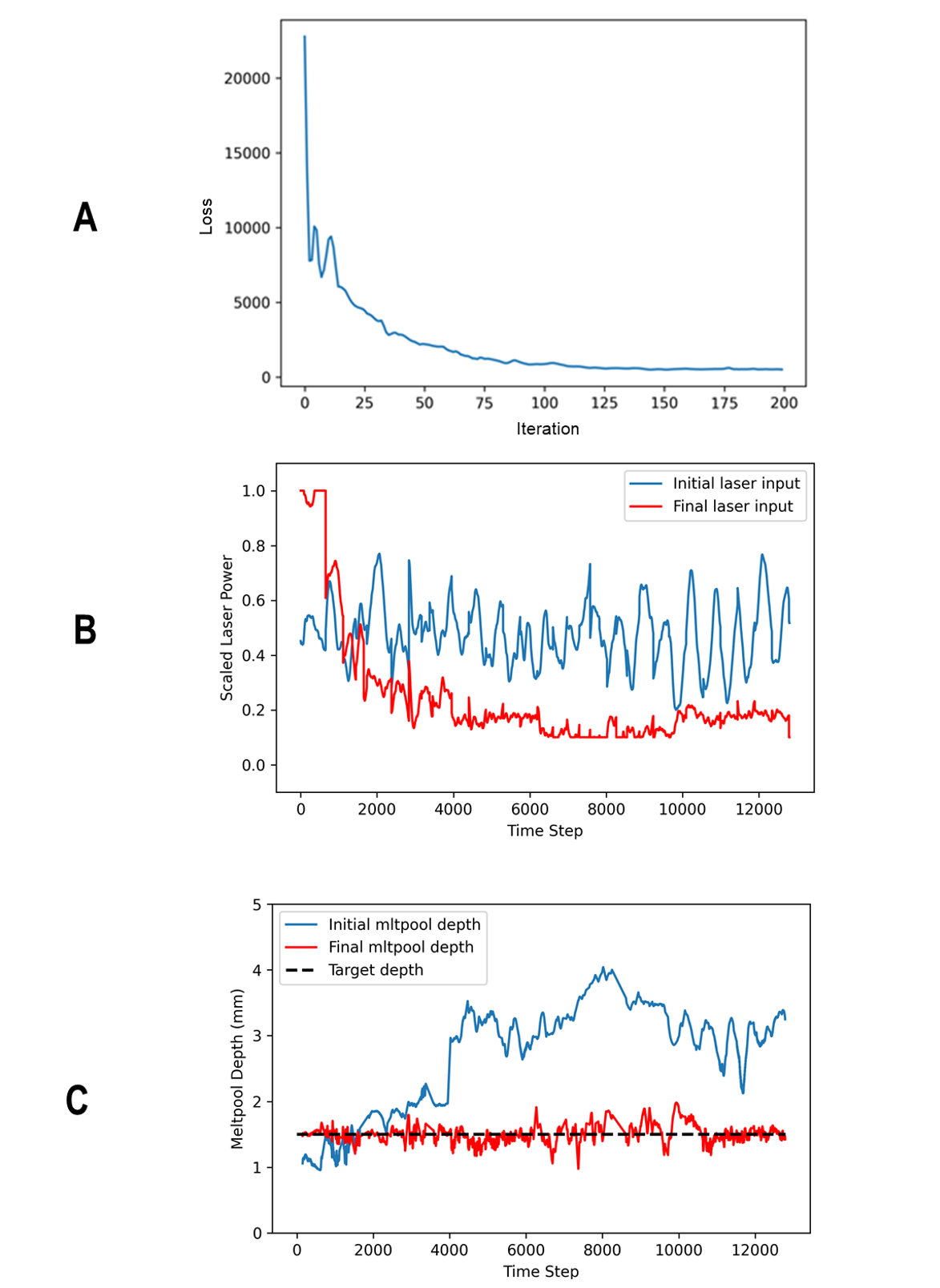}
	\caption{Optimization results for the third case study. (A) the evolution of MSE loss function between the desired melt pool depth and achieved depth. (B) the initial and final laser power after 200 optimization iterations on neural network parameters. (C) the initial melt pool depth, final depth after 200 optimization iterations, and target depth used in loss function definition.}
	\label{fig:fig10}
\end{figure}

Our results shown in Fig.\ref{fig:fig10} indicate that starting from a randomly initialized laser power pattern, we can learn a high-dimensional time-series laser power to control melt pool depth over thousands of FEM time steps. The evolution of MSE loss between desired and predicted melt pool depth is plotted in  Figure Fig.\ref{fig:fig10}A. Without optimization, we see a rapid increase in melt pool depth due to the heat accumulation especially halfway during the simulation as the laser builds the bottleneck of the hourglass geometry (see the blue curve in Figure Fig.\ref{fig:fig10}C). However, after optimization, the laser power sharply decreases after the first few lasers to keep melt pool depth close to the target depth and gradually increases it toward the end of the build to account for additional material deposition of the top layers of the hourglass geometry (see the red curves in Figure Fig.\ref{fig:fig10}B-C).

\section{Conclusions and Future Work}

In this article, we laid out our vision on differentiable physics-based simulation in manufacturing processes and particularly in AM. We demonstrated the capability of differentiable simulations to design and optimize the various process and material properties of the process in three representative case studies for (i) inferring static material and process parameters from partially observable data, (ii) designing time-series laser input to obtain a predefined thermal response, and (iii) designing time-series laser input to stabilize melt pool depth during the build. In all three cases, we showed that one can calculate the gradients using automatic differentiation and the gradients do not suffer from saturation or corruption even over tens of thousands of time steps. Therefore, the gradients can be effectively used in gradient-based optimization methods, such as Adam, to obtain favorable responses and eliminates the need for approximated ad-hoc solutions. This approach is particularly helpful in designing high-dimensional parameters, such as time-series parameters, where other optimization methods fail to provide a viable solution.

While we believe this approach shows great promise, many research avenues require further investigation. The first issue with the widespread application of differentiable simulations is that not all operations are inherently differentiable. For instance, we found it difficult to establish differentiable operations to perform a search and find the last time that a material undergoes the melting process. Note that one can pre-compute the step that the remelting happens and hard-code this information into a differentiable solution; however, developing a differentiable system to dynamically find this solution remains unsolved. Therefore, a main future research direction involves developing differentiable alternatives for many discontinuous algorithms and formulations in scientific computing. Finally, as differentiable simulation and optimization rely on local gradients, it is prone to stagnation in locally optimal solutions, and it can be heavily influenced by the initialization. Moreover, as in all general non-convex optimization methods, the solution is not unique. Therefore, careful design of the optimization process and initialization method is often needed to ensure satisfactory results.

\section{Acknowledgements}
This work was supported by the Vannevar Bush Faculty Fellowship N00014-19-1-2642, National Institute of Standards and Technology (NIST) – Center for Hierarchical Material Design (CHiMaD) under grant No.70NANB14H012, and the National Science Foundation (NSF) – Cyber-Physical Systems (CPS) under grant No.CPS/CMMI-1646592.

\bibliographystyle{unsrtnat}






\end{document}